\documentclass[twocolumn,english]{revtex4-2}
\usepackage[T1]{fontenc}
\usepackage[latin9]{inputenc}
\usepackage{color}
\usepackage{amsmath}
\usepackage{amssymb}
\usepackage{graphicx}

\makeatletter
\usepackage{physics}

\usepackage{babel}

\makeatother

\usepackage{babel}
\begin{document}
\title{Temporal cavity soliton interaction in passively mode-locked semiconductor
lasers}
\author{Andrei G. Vladimirov}
\affiliation{Weierstrass Institute, Mohrenstr. 39, 10117 Berlin, Germany}

\begin{abstract}
Weak interaction of temporal cavity solitons due to gain saturation and recovery  in a delay differential model of a long cavity semiconductor
laser is studied numerically and analytically using an asymptotic
approach. It is shown that in addition to the usual soliton repulsion leading
to a harmonic mode-locking regimes a soliton attraction is also possible
in a laser with nonzero linewidth enhancement factor. It is shown
numerically that this attraction can lead either to a pulse merging
or to a pulse bound state formation. 
\end{abstract}
\maketitle

\section{Introduction}

Temporal cavity solitons (TCSs) are short nonlinear optical pulses
generated by mode-locked lasers and optical microresonators and preserving
their shape in the course of propagation \cite{grelu2012dissipative,herr2014temporal,kippenberg2018dissipative}.
In lasers, unlike the usual self-starting mode-locked pulses
generated above the linear laser threshold, TCSs coexist with stable
laser off regime and require a finite perturbation for their excitation.
For example, when the cavity length of a laser with a semiconductor
gain medium is sufficiently large, usual mode-locked pulses can be
transformed into TCSs \cite{Marconi} corresponding to a non-self-starting
mode-locking regime. In many practical situations when more than one
TCSs are exited in an optical cavity weak interaction between the
TCSs may take place via their exponentially decaying tails. Spatial
and temporal dissipative soliton interaction in lasers with saturable
absorbers was studied in many publications in the case when the gain
and absorber populations were adiabatically eliminated and the interaction
took place only via the overlapping electric fields of the pulses
\cite{malomed1993bound,akhmediev1997multisoliton,akhmediev2001interaction,vladimirov2001stable,ablowitz2009soliton}.
Less investigated is the interaction of mode-locked pulses in the
presence of finite relaxation times of the gain and/or absorber media.
In this case the electromagnetic field saturates gain and absorption
behind the pulse and their slow relaxation can affect the position
of the next pulse traveling in the cavity. This type of interaction
was studied in Ref. \cite{kutz1998stabilized,soto1999multisoliton,nizette2006pulse,zaviyalov2012impact,camelin2016electrical}.
In particular, it was demonstrated theoretically and verified experimentally
with solid state and fiber lasers \cite{kutz1998stabilized} that
the interaction due to gain depletion and very slow recovery can produce
a repulsive force between adjacent pulses leading to the formation
of harmonic mode-locking regimes. Similar conclusion was made in Ref.
\cite{nizette2006pulse} using the delay differential equation (DDE)
model \cite{VT05,VTK,VT04} of a mode-locked monolithic semiconductor
laser, where similarly to Ref. \cite{kutz1998stabilized} the gain
recovery time was much longer than the cavity round trip time. Here using the same DDE model I consider the mode-locked pulse interaction in the TCS regime, where the cavity round trip time is sufficiently long, much longer than the gain recovery time. Basing on asymptotic approach the equations governing slow evolution of the time separation
and phase difference of the interacting TCSs are derived and analyzed.
Asymptotic study of weak TCS interaction in DDE models of optical
systems was already carried out earlier in \cite{puzyrev2017bound,munsberg2020topological,vladimirov2022short}.
However, only in Ref.~\cite{vladimirov2022short} devoted to the TCS interaction
in nonlinear mirror mode-locked laser and here a closed analytical
form of the interaction equations is derived. Using these equations
I show that the TCS interaction scenarios can be more rich than those
described in \cite{kutz1998stabilized,nizette2006pulse}. Apart from
the pulse repulsion resulting in a harmonic mode-locking regime, TCS
attraction leading ether to pulse merging or bound state formation can
take place in a laser with nonzero linewidth enhancement factor. Note,
that soliton attraction leading to a bound state formation was observed
earlier in Ref. \cite{soto1999multisoliton} in a complex Ginzburg-Landau
equation type mode-locked laser model with second order dispersion and in the
DDE model of a nonlinear mirror mode-locked laser \cite{vladimirov2022short}.
Note, however, that ulike the present work, in both those papers the
Kerr nonlinearity played a decisive role in the process of the pulse
formation. Furthermore, since Ref. \cite{soto1999multisoliton} considers
the limit of infinitely large gain recovery time, the mechanism of
the pulse interaction in this paper was different and can be attributed
to the saturation and slow recovery of the absorption, rather than
the gain. 

\section{Model equations}

The DDE model of a passively mode-locked semiconductor laser for the
electric field amplitude $A\left(t\right)$ at the entrance of the
laser absorber section, saturable gain $G\left(t\right)$, and saturable
absorption $Q\left(t\right)$ in the gain and absorber sections, respectively, can
be written in the form \cite{VT05,VTK,VT04}: 
\begin{equation}
\gamma^{-1}\partial_{t}A+\left(1+i\omega\right)A=R{(t-T)}A{(t-T)},\label{eq:DDE1}
\end{equation}
\begin{equation}
\partial_{t}G=g_{0}-\gamma_{g}G-e^{-Q}\left(e^{G}-1\right)|A|^{2},\label{eq:DDE2}
\end{equation}
\begin{equation}
\partial_{t}Q=q_{0}-\gamma_{q}Q-s\left(1-e^{-Q}\right)|A|^{2},\label{eq:DDE3}
\end{equation}
with 
\[
R{(t)}=\sqrt{\kappa}e^{(1-i\alpha_{g})G{(t)}/2-(1-i\alpha_{q})Q{(t)}/2+i\phi-i\omega T}.
\]
Here $t$ is the time variable, $\kappa$ is the attenuation factor
describing linear non-resonant intensity losses per cavity round trip,
$\alpha_{g}$ and $\alpha_{q}$ are the linewidth enhancement factors
in the gain and absorber sections, respectively. The time delay parameter
$T$ stands for the cold cavity round trip time, $\gamma$ is the
spectral filtering bandwidth, $\gamma_{g}$ and $\gamma_{q}$ are
the normalized carrier relaxation rates in the gain and absorber sections, and $s$ is the ratio of the saturation intensities in these sections. The pump parameter $g_{0}$ depends on the injection current in the gain section, while $q_{0}$ is the unsaturated
loss parameter, which depends on the inverse voltage applied to the
absorber section. The parameter $\phi$ is the phase shift describing
the detuning between the central frequency of the spectral filter
and the closest cavity mode and $\omega$ the reference frequency.

It is well known that in a certain parameter domain Eqs. (\ref{eq:DDE1})-(\ref{eq:DDE3})
demonstrate pulsed solutions corresponding to fundamental single pulse
and harmonic multipulse mode-locking regimes \cite{VT05,VTK,VT04}.
Furthermore, it was shown in \cite{Marconi} that when the laser
cavity is sufficiently long, so that the round trip time is much larger
than the gain relaxation time, these pulses can be transformed into
TCSs sitting on the stable laser off solution. In this situation two
well separated mode-locking pulses can interact only weakly via their
exponentially decaying tails. Furthermore, when the pulses are sufficiently
far away from one another, this interaction is mainly due to the gain
component $G$, which usually decays much slower than the electric
field envelope $A$ and the saturable absorption $Q$. Note, however,
that when the distance between the TCSs becomes small enough the interaction
via absorber component also might come into play and even lead to
the pulse bound state formation, see Ref. \cite{soto1999multisoliton},
where the case of infinitely large gain recovery time was considered.

In order to derive the TCS interaction equations we rewrite the model
equations in a more general real vector form 
\begin{equation}
\partial_{t}\mathbf{U}=\mathbf{F}_{\omega}\left(\mathbf{U}\right)+\mathbf{H}_{\omega}\left[\mathbf{U}{(t-T)}\right],\label{eq:general_form}
\end{equation}
where ${\bf U}=\left(\begin{array}{cccc}
U_{1} & U_{2} & U_{3} & U_{4}\end{array}\right)^{T}$ is real column vector with $U_{1}=\Re A$, $U_{2}=\Im A$, $U_{3}=G-g_{0}/\gamma_{g}$,
$U_{4}=Q-q_{0}/\gamma_{q}$, 
\[
\mathbf{F}_{\omega}\left(\mathbf{U}\right)=\left(\begin{array}{c}
-\gamma\left(U_{1}-\omega U_{2}\right)\\
-\gamma\left(U_{2}+\omega U_{1}\right)\\
-\gamma_{g}U_{3}-e^{-U_{4}-Q_{0}}\left(e^{U_{3}+G_{0}}-1\right)\left(U_{1}^{2}+U_{2}^{2}\right)\\
-\gamma_{q}U_{4}-s\left(1-e^{-U_{4}-Q_{0}}\right)\left(U_{1}^{2}+U_{2}^{2}\right)
\end{array}\right),
\]
and 
\[
\mathbf{H}_{\omega}\left(\mathbf{U}\right)=\left(\begin{array}{c}
-\Re \left[R\left(U_1+i U_2\right)\right]\\
-\Im \left[R\left(U_1+i U_2\right)\right]\\
0\\
0
\end{array}\right)
\]
with 
\[
R\left(t\right)=\gamma\sqrt{\kappa}e^{(1-i\alpha_{g})\left(U_{3}+g_{0}/\gamma_g\right)/2-(1-i\alpha_{q})\left(U_{4}+q_{0}/\gamma_q\right)/2-i\omega T}.
\]
.

\section{Temporal cavity soliton}

\begin{figure}
\begin{center}
\includegraphics[scale=0.5]{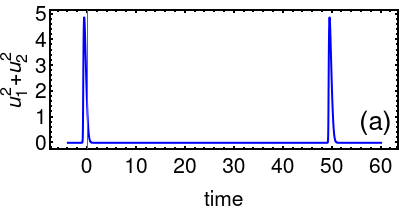}\\
\includegraphics[scale=0.5]{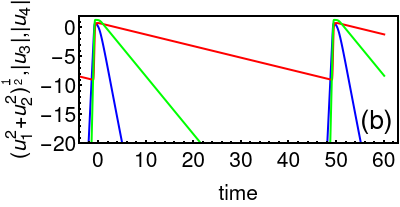}
\caption{Intensity time trace of the periodic TCS solution of Eqs. (\ref{eq:DDE1})-(\ref{eq:DDE3})
(a). Temporal evolution of the absolute values of the field envelope
(blue), gain (red) and loss (green) components of the TCS solution
in logarithmic scale (b). Parameters are: $\alpha_{g}=\alpha_{q}=0$, $g_{0}=0.5$,
$q_{0}=4.0$, $\kappa=0.8$, $s=10.0$, $\gamma=5.0$, $\gamma_{g}=0.2$, $\gamma_{q}=1.0$,
$T=50.0$. The solution period is $T_{0}=50.138425$ and $\omega_{0}=0$.\label{fig:Periodic-TCS-solution}}
\end{center}
\end{figure}

Let us assume that the inequalities 
\begin{equation}
\gamma^{-1}<\gamma_{q}^{-1}<\gamma_{g}^{-1}\ll T,\label{eq:ineq}
\end{equation}
for the relaxation rates in the model equations (\ref{eq:DDE1})-(\ref{eq:DDE3})
are satisfied. This means that the round trip time in a multimode semiconductor laser cavity is sufficiently long, much longer than the gain relaxation time. In
this case the DDE model can have TCS solutions \cite{Marconi}. We
will assume that such a solution corresponding to a narrow mode-locked
pulse with the duration $\tau_{p}\sim\gamma^{-1}$ exists in a certain
parameter domain and is given by $\omega=\omega_{0}$ and $\mathbf{U}=\mathbf{u}=\left(\begin{array}{cccc}
u_{1} & u_{2} & u_{3} & u_{4}\end{array}\right)^{T}$ in terms of Eq. (\ref{eq:general_form}). Here $\mathbf{u}\left(t\right)=\mathbf{u}\left(t+T_{0}\right)$
is periodic in time with the period $T_{0}$ close to the delay time
$T$. In terms of the original model equations (\ref{eq:DDE1})-(\ref{eq:DDE3})
we have ${\bf u}=\left[\begin{array}{cccc}
\Re A_{0}\left(t\right) & \Im A_{0}\left(t\right) & G_{0}\left(t\right)-g_{0}/\gamma_{g} & Q_{0}\left(t\right)-q_{0}/\gamma_{q}\end{array}\right]^{T}$, where $A_{0}\left(t\right)$, $G_{0}\left(t\right)$, and $Q_{0}\left(t\right)$
is a $T_{0}$-periodic TCS solution of these equations. Numerically
calculated intensity time trace of the TCS solution is shown in Fig.~\ref{fig:Periodic-TCS-solution}(a).

The decay rates of the TCS tails are determined by the following linearzation
\cite{vladimirov2019dynamics,yanchuk2019temporal,stohr2023temporal} of the model equations
(\ref{eq:DDE1})-(\ref{eq:DDE3}), on the trivial solution: 
\begin{equation}
\gamma^{-1}\partial_{t}a+\left(1+i\omega_{0}\right)a=R_{0}a\left(t+\delta\right),\label{eq:LinearA}
\end{equation}
\begin{equation}
\partial_{t}v_{3}=-\gamma_{g}v_{3}.\label{eq:LinearG}
\end{equation}
\begin{equation}
\partial_{t}v_{4}=-\gamma_{q}v_{4}.\label{eq:LinearQ}
\end{equation}
where $a=v_{1}+iv_{2}$, $\mathbf{v}=\left(\begin{array}{cccc}
v_{1} & v_{2} & v_{2} & v_{4}\end{array}\right)^{T}$ is a small perturbations vector, 
$$R_{0}=\sqrt{\kappa}e^{(1-i\alpha_{g})g_0/(2\gamma_g)-(1-i\alpha_{q})q_0/(2\gamma_q)+i\phi+i\omega_{0}\delta},$$
and the time advance parameter is $\delta=T_{0}-T$. It follows from
Eqs. (\ref{eq:LinearG}) and (\ref{eq:LinearQ}) that the decay rates
of the TCS gain and absorber components at large positive times $t$
are determined by the corresponding eigenvalues $\lambda_{g,q}=-\gamma_{g,q}$,
while Eq. (\ref{eq:LinearA}) has an infinite number of eigenvalues
defined by 
\begin{equation}
\lambda_{k}=-\gamma\left(1+i\omega_{0}\right)-\delta^{-1}W_{k}\left[-\gamma\delta e^{-\left(1+i\omega_{0}\right)\gamma\delta}R_{0}\right].\label{eq:lambert}
\end{equation}
where $W_{k}$ is the Lambert function with the index $k=0,\pm1,\pm2\dots$. For the parameter values of Fig.~\ref{fig:Periodic-TCS-solution} we get $\lambda_0=-3.741$ and $\lambda_{-1}=16.892$, while all other eigenvalues are complex and have positive real parts greater than $\lambda_{-1}$. 

Assuming that the origin of the time coordinate, $t=0$, is located
at the TCS power peak we get from (\ref{eq:LinearG})-(\ref{eq:lambert})
that for $t>0$ far away from the TCS core its trailing edge can be expressed as
\begin{equation}
u_{1,2}\sim b_{1,2}e^{\lambda_{0}t},\quad u_{3}\sim b_{3}e^{-\gamma_{g}t},\quad u_{4}\sim b_{4}e^{-\gamma_{q}t},\label{eq:TCS_asympt}
\end{equation}
where $b_{1,2,3,4}$ are real constants that can be calculated numerically
for given parameter values of Eqs. (\ref{eq:DDE1})-(\ref{eq:DDE3}).

Next, let us consider the leading tail of the TCS at negative
times, $t<0$. Since Eqs. (\ref{eq:LinearG}) and (\ref{eq:LinearQ}) have
no eigenvalues with positive real parts, gain and absorber components
of the TCS leading edge, $u_{3}$ and $u_{4}$ decay faster than exponentially
in negative time \cite{vladimirov2019dynamics}. The
field component $u_{1}+iu_{2}$ of the leading tail decays exponentially at t<0 
with the decay rate determined by the eigenvalue $\lambda_{-1}$ 
having smallest positive real part. Since the inequality $\gamma_{g,q}<|\lambda_{0}|<\lambda_{-1}$ is satisfied, the field component of the TCS decays
faster in both time directions than the gain and absorber components
in positive time. This means that the interaction via the electromagnetic
field component can be neglected when considering the interaction
of two well separated TCSs. Such type of interaction is typical of
lasers with slow gain and absorption and can be viewed as the long-range
interaction \cite{vladimirov2022short} unlike the short range interaction via overlapping electric fields considered in \cite{malomed1993bound,akhmediev1997multisoliton,akhmediev2001interaction,vladimirov2001stable,turaev2012long}.
Furthermore, since gain and absorber components of a TCS decay faster
than exponentially in negative time, the leading tails of the TCSs
can be neglected in the derivation of the interaction equations.  Figure~\ref{fig:Periodic-TCS-solution}(b)
shows the time evolution of the absolute values of field envelope
$\sqrt{U_{01}^{2}+U_{02}^{2}}$, gain $\left|U_{03}\right|$, and
absorption $\left|U_{04}\right|$ components of this solution in logarithmic
scale. It is seen from this figure that the gain component dominates
over the field and absorption ones during almost all the time interval
between the two consequent pulses.

\begin{figure}
\begin{center}
\includegraphics[scale=0.5]{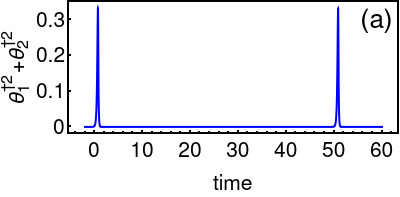}\\
\includegraphics[scale=0.5]{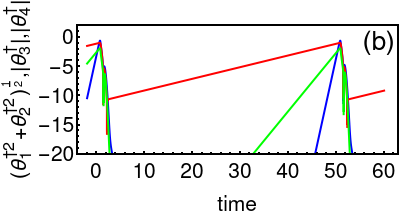}
\caption{Intensity time trace of the adjoint neutral mode $\boldsymbol{\theta}^{\dagger}$
as a function of time (a). Temporal evolution of the absolute values of the field (blue),
gain (red), and loss (green) components of the translational adjoint
neutral mode $\boldsymbol{\theta}^{\dagger}$ in logarithmic scale
(b). Parameters are the same as in Fig.~\ref{fig:Periodic-TCS-solution}.\label{fig:adjoint_mode}}
\end{center}
\end{figure}

Linear stability of the TCS is determined by linearizing Eq. (\ref{eq:general_form})
at the solution $\mathbf{U}=\mathbf{u}$ and calculating the spectrum
of the resulting linear operator ${\cal L}$. Due to the translational
and phase shift symmetries of the model equations (\ref{eq:DDE1})-(\ref{eq:DDE3}),
$\mathbf{U}\left(t\right)\to\mathbf{U}\left(t-t_{0}\right)$ and $U_{1}+iU_{2}\to\left(U_{1}+iU_{2}\right)e^{i\phi_{0}}$
with arbitrary constants $t_{0}$ and $\phi_{0}$, the operator ${\cal L}$
has a pair of zero eigenvalues corresponding to the neutral (Goldstone)
modes given by $\boldsymbol{\theta}=\partial_{t}{\bf u}$ and $\boldsymbol{\varphi}=\left(\begin{array}{cccc}
-u_{2} & u_{1} & 0 & 0\end{array}\right)^{T}$, respectively, ${\cal L}\boldsymbol{\theta}=-\partial_{t}\boldsymbol{\theta}+{\cal B}\left({\bf u}\right)\boldsymbol{\theta}+{\cal C}\left[{\bf u}\left(t-T\right)\right]\boldsymbol{\theta}^{\dagger}\left(t-T\right)=0$,
where ${\cal B}\left({\bf u}\right)$ {[}(${\cal C}\left({\bf u}\right)${]}
is the linearization matrix of $\mathbf{F}_{\omega_{0}}\left(\mathbf{U}\right)$
{[}$\mathbf{H}_{\omega_{0}}\left(\mathbf{U}\right)${]} at $\mathbf{U}=\mathbf{u}$
and ${\cal L}\boldsymbol{\varphi}=0$. Let us assume that the TCS
is stable, which means that the rest of the spectrum of the operator
${\cal L}$ lies in the left half of the complex plane. Similarly,
the linear operator ${\cal L}^{\dagger}$ adjoint to ${\cal L}$ has
a pair of zero eigenvalues associated with the so-called adjoint neutral
modes $\boldsymbol{\theta}^{\dagger}$ and $\boldsymbol{\varphi}^{\dagger}$,
${\cal L}^{\dagger}\boldsymbol{\theta}^{\dagger}=\partial_{t}\boldsymbol{\theta}^{\dagger}+\boldsymbol{\theta}^{\dagger}{\cal B}\left({\bf u}\right)+\boldsymbol{\theta}^{\dagger}\left(t+T\right){\cal C}\left({\bf u}\right)=0$
and ${\cal L}^{\dagger}\boldsymbol{\varphi}^{\dagger}=0$. The adjoint
neutral modes are assumed to be biorthogonal to the
neutral modes, $\left\langle \boldsymbol{\theta}^{\dagger}\cdot\boldsymbol{\varphi}\right\rangle =\left\langle \boldsymbol{\varphi}^{\dagger}\cdot\boldsymbol{\theta}\right\rangle =0$
and $\left\langle \boldsymbol{\theta}^{\dagger}\cdot\boldsymbol{\theta}\right\rangle =\left\langle \boldsymbol{\varphi}^{\dagger}\cdot\boldsymbol{\varphi}\right\rangle =1$,
where $\left\langle \boldsymbol{x}^{\dagger}\cdot\boldsymbol{y}\right\rangle =\int_{0}^{\tau_{0}}\boldsymbol{x}^{\dagger}\cdot\boldsymbol{y}dt$.
Since the adjoint operator ${\cal L}^{\dagger}$ is obtained from
${\cal L}$ by the transformations including the time reversal, $t\to-t$,
the asymptotic behavior of the row vector adjoint neutral modes $\boldsymbol{\theta}^{\dagger}=\left(\begin{array}{cccc}
\theta_{1}^{\dagger} & \theta_{2}^{\dagger} & \theta_{3}^{\dagger} & \theta_{4}^{\dagger}\end{array}\right)$ and $\boldsymbol{\varphi}^{\dagger}=\left(\begin{array}{cccc}
\varphi_{1}^{\dagger} & \varphi_{2}^{\dagger} & \varphi_{3}^{\dagger} & \varphi_{4}^{\dagger}\end{array}\right)$ at sufficiently large negative times $t<0$ is given by 
\begin{equation}
\theta_{1,2}^{\dagger}\sim c_{1,2}e^{-\lambda_{0}t},\quad\theta_{3}^{\dagger}\sim c_{3}e^{\gamma_{g}t},\quad\theta_{4}^{\dagger}\sim c_{4}e^{\gamma_{q}t},\label{eq:adjoint_asympt}
\end{equation}
\begin{equation}
\varphi_{1,2}^{\dagger}\sim d_{1,2}e^{-\lambda_{0}t},\quad\varphi_{3}^{\dagger}\sim d_{3}e^{\gamma_{g}t},\quad\varphi_{4}^{\dagger}\sim d_{4}e^{\gamma_{q}t},\label{eq:adjoint_asympt-1}
\end{equation}
where $c_{1,2,3,4}$ and $d_{1,2,3,4}$ are real coefficients, which
can be calculated numerically. Similarly to the leading tail of the
TCS solution, the trailing tail of the gain and absorber components
of the adjoint neutral modes decay faster than exponentially at $t>0$. Therefore, trailing tails of the adjoint neutral modes will
be neglected when deriving the TCS interaction equations. The temporal
evolution of the field, gain, and loss components of the translational
adjoint neutral mode $\boldsymbol{\theta}^{\dagger}=\left(\begin{array}{cccc}
\theta_{1}^{\dagger} & \theta_{2}^{\dagger} & \theta_{3}^{\dagger} & \theta_{4}^{\dagger}\end{array}\right)$ are shown in Fig.~\ref{fig:adjoint_mode}. We see that similarly
to Fig.~\ref{fig:Periodic-TCS-solution} the gain component $\theta_{3}^{\dagger}$
of the adjoint neutral mode dominates almost everywhere between the
consequent mode-locked pulses. Therefore, one can conclude that the
pulse interaction via the field and absorber components can be neglected
for the parameter values of these figures.

\section{Interaction equations}

To derive the equations describing slow evolution of the time coordinates
and phases of weakly interacting TCSs we look for the solution of
Eq. (\ref{eq:general_form}) in the form of a sum of two unperturbed
TCS solutions plus a small correction ${\bf w(t)={\cal O}}\left(\epsilon\right)$
due to the interaction: 
\begin{eqnarray}
{\bf U} & = & \sum_{k=1}^{2}{\bf u}_{k}+{\bf w},\label{eq:Anzatz}
\end{eqnarray}
where ${\bf u}_{k}=\left(\begin{array}{cccc}
u_{1k} & u_{2k} & u_{3k} & u_{4k}\end{array}\right)^{T}$ with $u_{1k}+iu_{2k}=\left[u_{1}\left(t-\tau_{k}\right)+iu_{2}\left(t-\tau_{k}\right)\right]e^{-i\phi_{k}}$,
$u_{3k}=u_{3}\left(t-\tau_{k}\right)$, and $u_{4k}=u_{4}\left(t-\tau_{k}\right)$.
Coordinates $\tau_{k}$ and phases $\phi_{k}$ of the interacting
TCSs are assumed to be slow functions of time, $\partial_{t}\tau_{k},\partial_{t}\phi_{k}={\cal O}\left(\epsilon\right)$,
$k=1,2$. The small parameter $\epsilon$ characterizes weak overlap
of the TCSs. Similarly to the case of dissipative soliton interaction
in partial differential equation laser models \cite{vladimirov2001stable,turaev2012long,vladimirov2018effect,vladimirov2021dissipative},
the right hand side of the interaction equations obtained for our
DDE model can be expressed in terms of the TCS solutions and their
adjoint neutral modes evaluated at the point between the two TCSs
\cite{vladimirov2022short}. The details of the calculations are
given in the Appendix \ref{sec:Derivation}, where it is shown that
the interaction equations for the time separation $\Delta\tau=\tau_{2}-\tau_{1}$
and phase difference $\Delta\phi=\phi_{2}-\phi_{1}$ of a pair of
interacting $T_{0}$-periodic TCS take the form 
\begin{equation}
\partial_{t}\Delta\tau\approx\boldsymbol{\theta}_{1}^{\dagger}\left(T_{0}/2\right)\mathbf{u}_{2}\left(T_{0}/2\right)-\boldsymbol{\theta}_{2}^{\dagger}\left(0\right)\mathbf{u}_{1}\left(0\right),\label{eq:Int1-1}
\end{equation}
\begin{equation}
\partial_{t}\Delta\phi\approx\boldsymbol{\varphi}_{1}^{\dagger}\left(T_{0}/2\right)\mathbf{u}_{2}\left(T_{0}/2\right)-\boldsymbol{\varphi}_{2}^{\dagger}\left(0\right)\mathbf{u}_{1}\left(0\right),\label{eq:Int2-1}
\end{equation}
where $\boldsymbol{\theta}_{k}^{\dagger}=\boldsymbol{\theta}^{\dagger}(t-t_{k})$ and $\boldsymbol{\varphi}_{k}^{\dagger}=\boldsymbol{\varphi}^{\dagger}(t-t_{k})$
are the adjoint neutral modes evaluated at the $k$th TCS ($k=1,2$)
and without the loss of generality one can assume that $t=0$ and
$t=T_{0}/2$ correspond, respectively, to the middle point between
the two interacting o TCSs and the opposite point on a circle with
the circumference $T_{0}$. 

Substituting asymptotic expressions (\ref{eq:TCS_asympt}), (\ref{eq:adjoint_asympt}),
and (\ref{eq:adjoint_asympt-1}) into the interaction equations (\ref{eq:Int1-1})
and (\ref{eq:Int2-1}) and neglecting the field components gives 
\begin{eqnarray}
\partial_{t}\Delta\tau & = & K_{\tau g}\left[e^{-\gamma_{g}\left(T_{0}-\Delta\tau\right)}-e^{-\gamma_{g}\Delta\tau}\right]\\
 & + & K_{\tau q}\left[e^{-\gamma_{q}\left(T_{0}-\Delta\tau\right)}-e^{-\gamma_{q}\Delta\tau}\right],\label{eq:interact1}
\end{eqnarray}
\begin{eqnarray}
\partial_{t}\Delta\phi & = & K_{\phi g}\left[e^{-\gamma_{g}\left(T_{0}-\Delta\tau\right)}-e^{-\gamma_{g}\Delta\tau}\right]\\
 & + & K_{\phi q}\left[e^{-\gamma_{q}\left(T_{0}-\Delta\tau\right)}-e^{-\gamma_{q}\Delta\tau}\right]\label{eq:interact2}
\end{eqnarray}
with $K_{\tau g}=b_{3}c_{3}$, $K_{\tau q}=b_{4}c_{4}$, $K_{\phi g}=b_{3}d_{3}$,
and $K_{\phi q}=b_{4}d_{4}$. Interaction equations (\ref{eq:interact1})
and (\ref{eq:interact2}) describe the long-range interaction of two
well separated TCS via their gain and absorber components and do not
take into account the short range interaction via weakly overlapping
electric field envelopes of the TCSs. They reflect the fact that in
a ring cavity the interaction of the two TCSs is twofold. The trailing tail of the
first (second) TCS overlaps with the leading tail of the adjoint neutral
mode of the second (first) TCS which is located by $\Delta\tau$ ($T_{0}-\Delta\tau$)
behind it. This is reflected by the presence of the two exponential
terms in the square brackets of Eqs. (\ref{eq:interact1}) and (\ref{eq:interact2}).
As it was already noted above, due to the inequality $\gamma_{q}>\gamma_{g}$
typical of semiconductor lasers, the interaction force related to
the absorber component decays much faster than that due to the gain
component and the terms proportional to $K_{\tau q}$ and $K_{\phi q}$
can be neglected in the interaction equations. In the case of TCS
repulsion ($K_{\tau g}<0$) such type of twofold interaction leads
to a regime with two equally spaced pulses per cavity round trip corresponding
to a harmonic mode-locking regime.

\section{Results of numerical simulations}

\begin{figure}
\begin{center}
\includegraphics[scale=0.35]{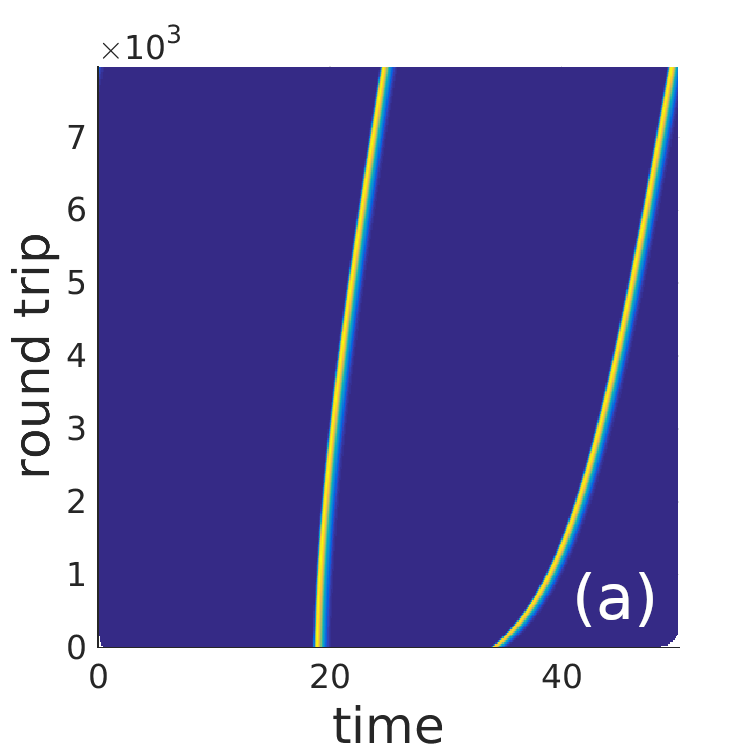}\includegraphics[scale=0.365]{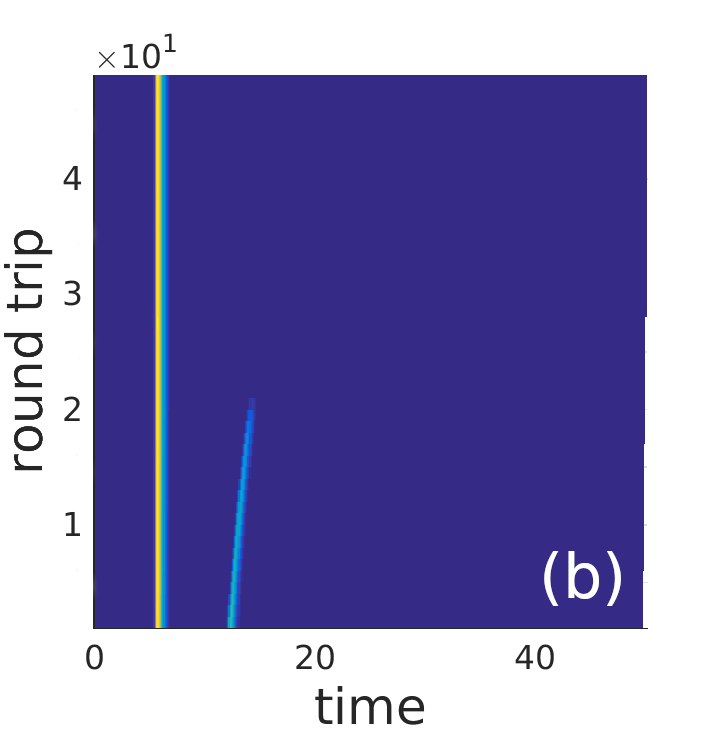}\\
 \includegraphics[scale=0.35]{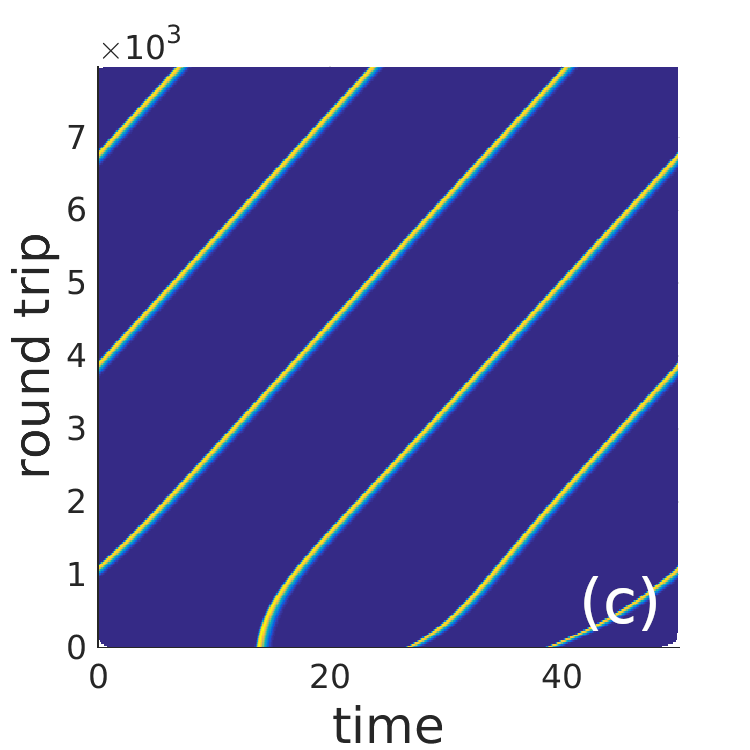}\includegraphics[scale=0.35]{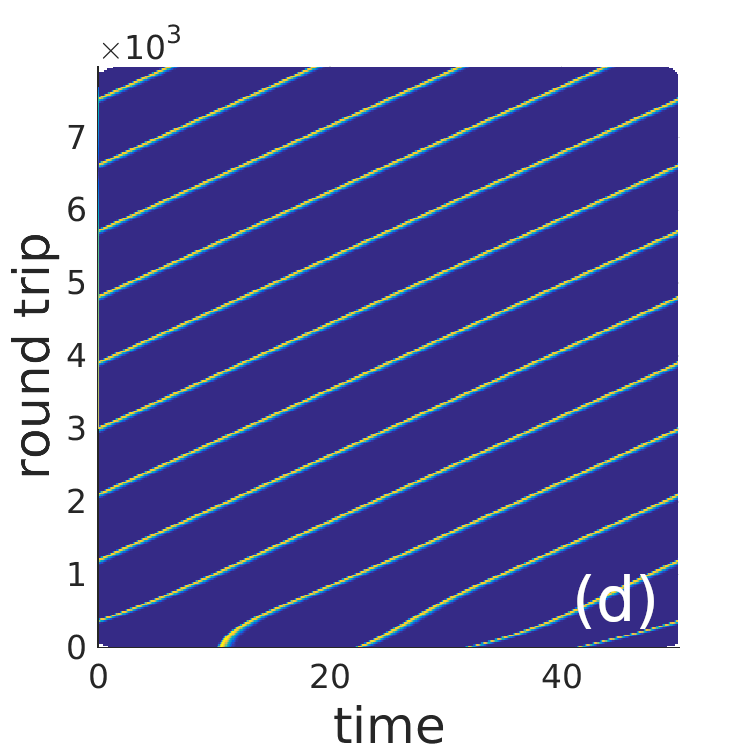}
\caption{TSC repulsion due to the interaction via the gain component leading to a harmonic mode-locking
regimes with two (a) and three (c) and four (d) pulses per cavity
round trip. Panel (b) illustrates repulsive interaction resulting
in the annihilation of the second pulse. (a), (b), and (c) - $g_{0}=0.5$.
(d) - $g_{0}=0.8$. Other parameters the same as in Fig.~\ref{fig:Periodic-TCS-solution}.
\label{fig:TSC-repulsion}}
\end{center}
\end{figure}

For the parameter values of Figs. \ref{fig:Periodic-TCS-solution}
and \ref{fig:adjoint_mode} corresponding to zero linewidth enhancement
factors, $\alpha_{g}=\alpha_{q}=0$, numerically we get $K_{\tau g}=-1.120$
and $K_{\tau q}=2.145$ in Eq. (\ref{eq:interact1}), while the second
interaction equation (\ref{eq:interact2}) transforms into $\partial_{t}\Delta\phi=0$
due to the relation $K_{\phi g}=K_{\phi q}=d_{3}=d_{4}=0$, which
is the consequence of $\omega_{0}=\Im A_{0}=0$. Negative value of
$K_{\tau g}$ means that the TCS interaction is repulsive, while positive
$K_{\tau q}$ corresponds to TCS attraction via the absorber component.
For the parameter values of these figures, however, the interaction
via gain component dominates for almost all sufficiently large soliton
separations, as it was discussed above, and the soliton attraction
due to the absorber component is hardly possible to observe. This
is illustrated in Fig.~\ref{fig:TSC-repulsion} where the soliton
repulsion is illustrated by numerical integration of Eqs. (\ref{eq:DDE1})-(\ref{eq:DDE3})
using the RADAR5 code \cite{guglielmi2001implementing}. The initial
condition was taken as a sum of two or more well separated unperturbed
TCSs. Figure \ref{fig:TSC-repulsion}(a) shows the standard mechanism
of the harmonic mode-locking regime formation as a result of the repulsion
of a pair of TCSs due to the interaction via the gain component. Figure
\ref{fig:TSC-repulsion}(b) was obtained for the same parameter values
but with smaller initial separation of the two TCSs. It is seen that
during the first stage of the interaction there is still repulsion
between the TCSs, but later the second TCS loses energy and disappears.
The equation $\partial_{t}\Delta\phi=0$ means that the TCS phase
difference remains almost constant in the course of the interaction.
This difference is affected only by a very weak overlap of the field
componetns which are neglected in the derivation of the interaction
equations (\ref{eq:interact1}) and (\ref{eq:interact2}). Repulsive interaction
of three and four TCSs leading to the development of harmonic mode-locking
regimes with three and four pulses per cavity round trip are illustrated in Fig.~\ref{fig:TSC-repulsion}(c)
and \ref{fig:TSC-repulsion}(d), respectively.

\begin{figure}
\includegraphics[scale=0.40]{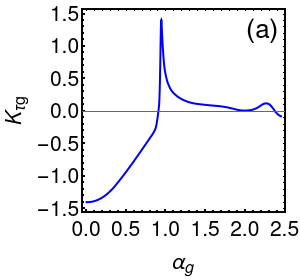}\includegraphics[scale=0.40]{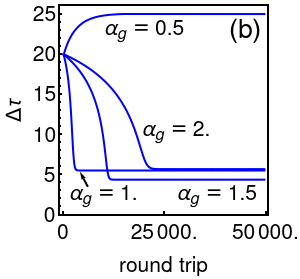}\caption{Interaction coefficient $K_{\tau g}$ as a function of $\alpha_{g}$
(a) and pulse time separation as a function of the round trip number
(b). $g_{0}=0.8.$ Other parameters are the same as in Fig.~\ref{fig:Periodic-TCS-solution}\label{fig:Coefficient}}
\end{figure}

\begin{figure}
\begin{center}
\includegraphics[scale=0.35]{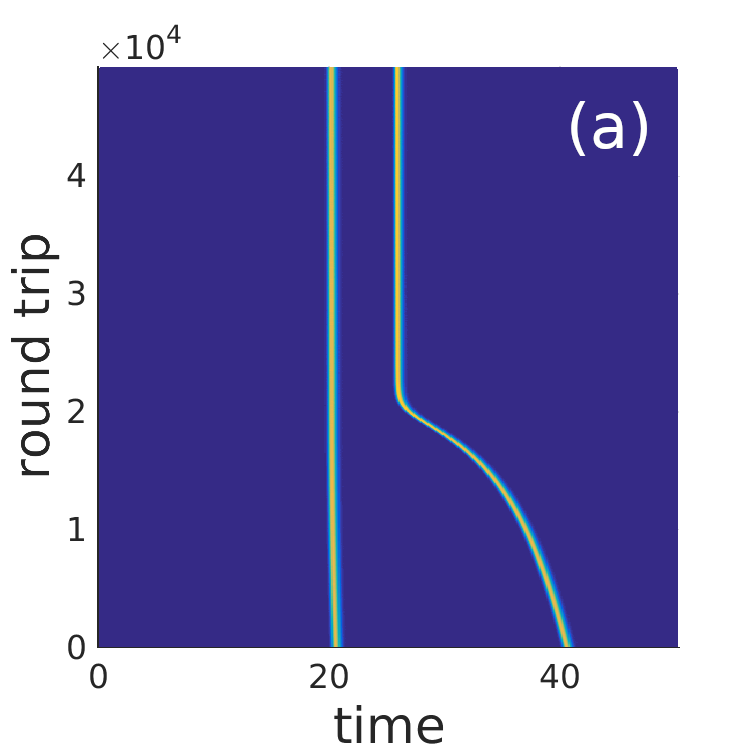}\includegraphics[scale=0.35]{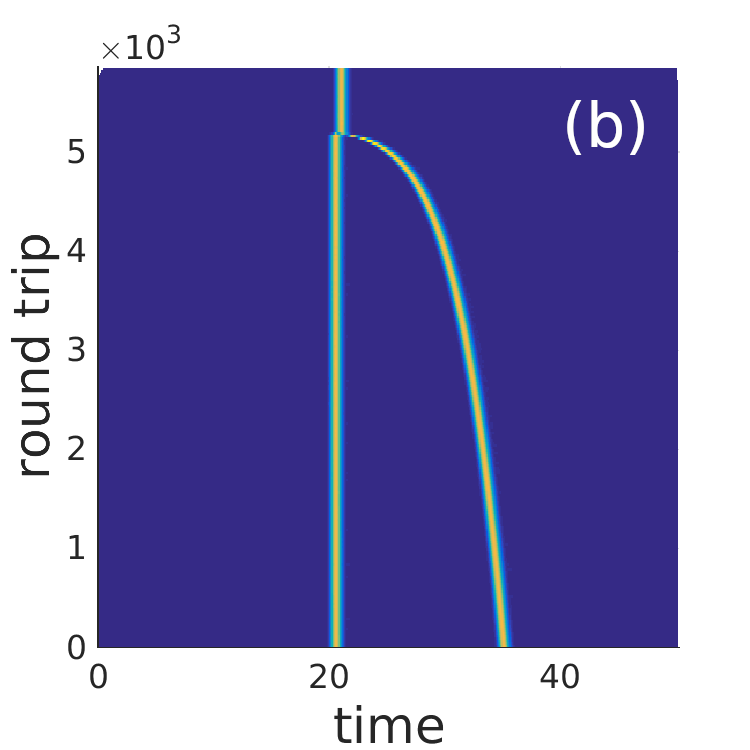}\caption{TCS interaction resulting in pulse bound state formation at $q_{0}=5.0$
(a) and pulse merging at $q_{0}=4.0$ (b). $g_{0}=0.8$,$\text{\ensuremath{\alpha_{g}=2.0}}$,
$\alpha_{q}=0$. Other parameters are as in Fig.~\ref{fig:TSC-repulsion}\label{fig:TCs-interaction}}
\end{center}
\end{figure}

\begin{figure}
\begin{center}
\includegraphics[scale=0.40]{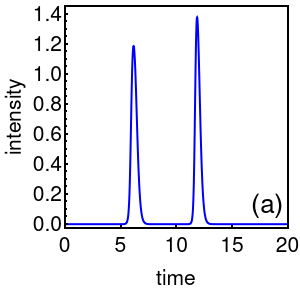}\includegraphics[scale=0.40]{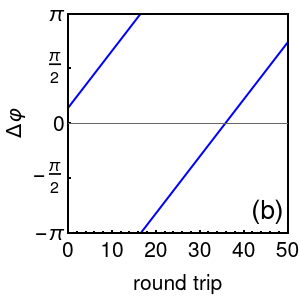}

\caption{Intensity time-trace (a) and pulse phase difference (b) of the TCS
bound state. Parameters are the same as in Fig.~\ref{fig:TCs-interaction}
(a). \label{fig:Phase-difference}}
\end{center}
\end{figure}

The dependence of the interaction coefficient $K_{\tau g}$ on the
linewidth enhancement factor $\alpha_{g}$ in the gain section is
shown in Fig.~\ref{fig:Coefficient}(a). It is seen that this dependence
is non-monotonous and has a pronounced resonant character. The interaction
coefficient is negative (TCS repulsion) when the linewidth enhancement
factor is sufficiently small, and it becomes positive (TCS attraction)
with the increase of $\alpha_{g}$ showing a sharp peak around $\alpha_{g}\approx0.94$.
Further increase of the $\alpha_{g}$ leads to a non-monotonous gradual
decrease of the interaction coefficient which becomes negative again
at $\alpha_{g}\gtrapprox2.37$. Numerical simulation of the TCS interaction
of Eqs. (\ref{eq:DDE1})-(\ref{eq:DDE3}) with $\alpha_{g}=2.0$,
which corresponds to a small positive values of the interaction coefficient, is
illustrated in Fig.~\ref{fig:TCs-interaction}. It is seen that the
interaction is very asymmetric, see Refs. \cite{camelin2016electrical,vladimirov2018effect,vladimirov2022short}
and Appendix \ref{sec:Derivation}. Figure \ref{fig:TCs-interaction}(b)
corresponding to $q_{0}=4.0$ and positive $K_{\tau g}\approx0.854\times10^{-2}$
shows the TCS attraction leading to the merging of two pulses when
one of them is annihilated after the collision. In Fig.~\ref{fig:TCs-interaction}(a)
obtained for $q_{0}=5.0$ and $K_{\tau g}\approx1.076\times10^{-2}$
the soliton attraction leads to a formation of a pulse bound state.
Since for $\alpha_{g}\neq0$ the relation $d_{3}=d_{4}=0$ does not
hold any more the TCS phases are evolving with round trip number in
the course of interaction. Therefore, the bound state shown in Fig.
\ref{fig:TCs-interaction}(a) is similar to the ``incoherent'' bound
state described in \cite{vladimirov2022short} with the phase difference
$\Delta\phi$ between two pulses growing monotonously in time, see
Fig.~\ref{fig:Phase-difference} illustrating the intensity time trace
and the evolution of the TCS phase difference of the incoherent bound
state. It was demonstrated in \cite{vladimirov2022short} that due
to the electric field overlap of the interacting TCS such type of
bound states is characterized by slightly oscillating time separation
$\Delta\tau$. However, since the interaction via electric fields
is extremely small for the bound state shown in Fig.~\ref{fig:TCs-interaction}(a)
such oscillation is hardy possible to detect. Figure \ref{fig:Coefficient}(b)
shows the evolution of the inter-soliton time separation $\Delta\tau$
as a function of the round trip number obtained by direct numerical
simulation of the laser model (\ref{eq:DDE1})-(\ref{eq:DDE3}). The
parameter values are the same as in Fig.~\ref{fig:Coefficient}(a).
It is seen that for $\alpha_{g}=0.5$ when the interaction coefficient
$K_{\tau g}$ is negative the TCS interaction is repulsive leading
to a harmonic mode-locking regime. On the contrary, for $\alpha_{g}=1.0,1.5,2.0$,
which correspond to $K_{\tau g}>0$, the interaction results in the
formation of a pulse bound states. Furthermore, comparing Fig.~\ref{fig:Coefficient}(b)
with Fig.~\ref{fig:Coefficient}(a) we see that the smaller the interaction
coefficient the weaker is the interaction force and, hence, the longer
is the transient time before the equilibrium inter-pulse time separation
is achieved. The final inter-pulse distance in the bound state is,
however, only weakly dependent on the $\alpha_{g}$ and $K_{\tau g}$.
Note, that the time separation of the pulses in the incoherent bound
state shown in Fig.~\ref{fig:TCs-interaction}(a) is of the same order
of magnitude as the gain relaxation time. This is why \ref{fig:Phase-difference}(a)
the pulses in this bound state have significantly different peak powers
{[}see Fig.~\ref{fig:Phase-difference}(a){]} and cannot any more
be considered as individual TCSs. Therefore, the interaction equations
(\ref{eq:interact1}) and (\ref{eq:interact2}) are not valid any
more when the pulses are so close to one another. Indeed, in order
the bound state to be formed, the attraction predicted by the interaction
equations should be compensated by a repulsion at sufficiently small
inter-pulse distances. This repulsion dominating at small pulse separations
might be related to that in a laser with the cavity
round trip time shorter or much shorter than the gain relaxation time discussed in \cite{kutz1998stabilized,nizette2006pulse}. 

\section{Conclusion}

To conclude, using the DDE model interaction of two well separated TCSs
in a long cavity mode-locked semiconductor laser was studied numerically
and analytically. Interaction equations governing the slow evolution
of the time separation and phase difference of the TCSs were derived
and analyzed. Analytical results were compared to direct numerical
simulations of the DDE mode-locking model. It was demonstrated that
in addition to usual pulse repulsion predicted in \cite{kutz1998stabilized,nizette2006pulse}
an attractive TCS interaction is also possible in a laser with nonzero
linewidth enhancement factor. This attractive interaction can result
either in pulse merging or in a formation of incoherent pulse bound
state. In the latter case the repulsion force counteracting the soliton
attraction might be attributed to the standard mechanism of the mode-locking
pulse repulsion described in \cite{kutz1998stabilized,nizette2006pulse},
which acts beyond the TCS limit. Incoherent bound pulse state discussed
here is similar to that observed experimentally \cite{kokhanovskiy2020single}
and described theoretically \cite{vladimirov2022short} in a nonlinear
mirror mode-locked laser. It also has a similarity to the ``type
A'' pulse bound states reported in \cite{soto1999multisoliton}.
The mechanism of the latter bound states formation is, however, different
from that described here and can be related to the TCS attraction
due to the interaction via absorber component of the pulsed solution
in a laser with infinitely large gain relaxation time.

\begin{acknowledgments}
The support by the Deutsche Forschungsgemeinschaft (DFG projects No. 445430311 and No. 491234846) is gratefully acknowledged.  
\end{acknowledgments}

\appendix
\section{Derivation of the interaction equations\label{sec:Derivation}}

Substituting Eq. (\ref{eq:Anzatz}) into Eq. (\ref{eq:general_form}),
collecting the first order terms in small parameter $\epsilon$, and
applying solvability conditions \cite{halanay1966differential} to
the resulting equation yield 
\begin{gather}
\partial_{t}\tau_{k}=-\left\langle \boldsymbol{\theta}_{k}^{\dagger}\cdot{\bf P}\right\rangle ,\quad\partial_{t}\phi_{k}=-\left\langle \boldsymbol{\varphi}_{k}^{\dagger}\cdot{\bf P}\right\rangle ,\label{eq:Int1_tau_phi}\\
{\bf P=}-\partial_{t}\mathbf{u}_{\Sigma}+{\bf F}_{\omega_{0}}\left({\bf u}_{\Sigma}\right)+{\bf H}_{\omega_{0}}\left[{\bf u}_{\Sigma}\left(t-T\right)\right],\label{eq:P}
\end{gather}
where $\mathbf{u}_{\Sigma}=\mathbf{u}_{1}+\mathbf{u}_{2}$, $\left\langle \cdot\right\rangle =\int_{0}^{T_{0}}\cdot dt$,
and $\boldsymbol{\theta}_{k}^{\dagger}$ ($\boldsymbol{\varphi}_{k}^{\dagger}$)
is the adjoint translational (phase) neutral mode evaluated the $k$th
TCS, $k=1,2$.

Since ${\bf u}_{k}$ is the solution of Eq. (\ref{eq:general_form})
the equality $\sum_{k=1}^{2}\left\{ -\partial_{t}\mathbf{u}_{k}+{\bf F}_{\omega_{0}}\left({\bf u}_{k}\right)+{\bf H}_{\omega_{0}}\left[{\bf u}_{k}\left(t-T\right)\right]\right\} =0$
is satisfied. Subtracting this equality from (\ref{eq:P}) we get
\begin{gather}
\mathbf{P}={\bf F}_{\omega_{0}}\left({\bf u}_{\Sigma}\right)-\sum_{k=1}^{2}{\bf F}_{\omega_{0}}\left({\bf u}_{k}\right)+{\bf H}_{\omega_{0}}\left[{\bf u}_{\Sigma}\left(t-\tau\right)\right]\nonumber\\
-\sum_{k=1}^{2}{\bf H}_{\omega_{0}}\left[{\bf u}_{k}\left(t-T\right)\right].
\end{gather}
Therefore, the equation for $\tau_{2}$ in (\ref{eq:Int1_tau_phi})
is 
\begin{gather}
\partial_{t}\tau_{2}=-\left\langle \boldsymbol{\theta}_{2}^{\dagger}\cdot{\bf P}\right\rangle =-\left\langle \boldsymbol{\theta}_{2}^{\dagger}\cdot\left\{ {\bf F}_{\omega_{0}}\left({\bf u}_{\Sigma}\right)-\sum_{k=1}^{2}{\bf F}_{\omega_{0}}\left({\bf u}_{k}\right)\right.\right.\nonumber \\
+\left.\left.{\bf H}_{\omega_{0}}\left[{\bf u}_{\Sigma}\left(t-T\right)\right]-\sum_{k=1}^{2}{\bf H}_{\omega_{0}}\left[{\bf u}_{k}\left(t-T\right)\right]\right\} \right\rangle .\label{eq:tau2_1}
\end{gather}
Using $T_{0}-$periodicity of $\boldsymbol{\theta}_{2}^{\dagger}$
and ${\bf u}_{1,2}$ Eq. (\ref{eq:tau2_1}) can be rewritten as
\begin{gather}
\partial_{t}\tau_{2}=-\left\langle \boldsymbol{\theta}_{2}^{\dagger}\cdot\left[{\bf F}_{\omega_{0}}\left({\bf u}_{\Sigma}\right)-\sum_{k=1}^{2}{\bf F}_{\omega_{0}}\left({\bf u}_{k}\right)\right]\right\rangle \nonumber \\
-\left\langle \boldsymbol{\theta}_{2}^{\dagger}\left(t+T\right)\cdot\left[{\bf H}_{\omega_{0}}\left({\bf u}_{\Sigma}\right)-\sum_{k=1}^{2}{\bf H}_{\omega_{0}}\left({\bf u}_{k}\right)\right]\right\rangle .
\end{gather}

Next we split the integral $\left\langle \cdot\right\rangle =\int_{0}^{T_{0}}\cdot dt$
into two parts $\left\langle \cdot\right\rangle =\left\langle \cdot\right\rangle _{1}+\left\langle \cdot\right\rangle _{2}$,
where $\left\langle \cdot\right\rangle _{1}=\int_{-T_{0}/2}^{0}\cdot dt$
and $\left\langle \cdot\right\rangle _{2}=\int_{0}^{T_{0}/2}\cdot dt$
are the integrals over the intervals $\left[-T_{0}/2,0\right]$ and
$\left[0,T_{0}/2\right]$, respectively: 
\begin{eqnarray*}
\partial_{t}\tau_{2} & = & -\sum_{j=1}^{2}\left\langle \boldsymbol{\theta}_{2}^{\dagger}\cdot\left[{\bf F}_{\omega_{0}}\left({\bf u}_{\Sigma}\right)-\sum_{k=1}^{2}{\bf F}_{\omega_{0}}\left({\bf u}_{k}\right)\right]\right\rangle _{j}\\
 & - & \sum_{j=1}^{2}\left\langle \boldsymbol{\theta}_{2}^{\dagger}\left(t+T\right)\cdot\left[{\bf H}_{\omega_{0}}\left({\bf u}_{\Sigma}\right)-\sum_{k=1}^{2}{\bf H}_{\omega_{0}}\left({\bf u}_{k}\right)\right]\right\rangle _{j}.
\end{eqnarray*}
On the first interval $\left[-T_{0}/2,0\right]$, where $\mathbf{u}_{2}$
is small, one obtains 
\begin{eqnarray*}
{\bf F}_{\omega_{0}}\left({\bf u}_{\Sigma}\right)-{\bf F}_{\omega_{0}}\left({\bf u}_{1}\right) & \approx & {\cal B}_{1}\mathbf{u}_{2},\\
{\bf H}_{\omega_{0}}\left({\bf u}_{\Sigma}\right)-{\bf H}_{\omega_{0}}\left({\bf u}_{1}\right) & \approx & {\cal C}_{1}\mathbf{u}_{2},
\end{eqnarray*}
and
\[
{\bf F}_{\omega_{0}}\left({\bf u}_{2}\right)\approx{\cal B}_{0}\mathbf{u}_{2},\quad{\bf H}_{\omega_{0}}\left({\bf u}_{2}\right)\approx{\cal C}_{0}\mathbf{u}_{2}
\]
with ${\cal B}_{1}={\cal B}\left({\bf u}_{1}\right)$, ${\cal C}_{1}={\cal C}\left({\bf u}_{1}\right)$
{[}${\cal B}_{0}={\cal B}\left(0\right)$, and ${\cal C}_{0}={\cal C}\left(0\right)${]}
are the linearization matrices of ${\bf F}_{\omega_{0}}\left({\bf U}\right)$
and ${\bf H}_{\omega_{0}}\left({\bf U}\right)$ at ${\bf U}={\bf u}_{1}$
(${\bf U}=0$). Similarly, on the second interval $\left[0,T_{0}/2\right]$,
where $\mathbf{u}_{1}$ is small, one gets 
\begin{eqnarray*}
{\bf F}_{\omega_{0}}\left({\bf u}_{\Sigma}\right)-{\bf F}_{\omega_{0}}\left({\bf u}_{2}\right) & \approx & {\cal B}_{2}\mathbf{u}_{1},\\
{\bf H}_{\omega_{0}}\left({\bf u}_{\Sigma}\right)-{\bf H}_{\omega_{0}}\left({\bf u}_{1}\right) & \approx & {\cal C}_{2}\mathbf{u}_{1},
\end{eqnarray*}
where ${\cal B}_{2}={\cal B}\left({\bf u}_{2}\right)$ and ${\cal C}_{2}={\cal C}\left({\bf u}_{2}\right)$
and
\begin{equation}
{\bf F}_{\omega_{0}}\left({\bf u}_{1}\right)\approx{\cal B}_{0}\mathbf{u}_{1},\quad{\bf H}_{\omega_{0}}\left({\bf u}_{2}\right)\approx{\cal C}_{0}\mathbf{u}_{1}.\label{eq:aprox}
\end{equation}
Hence, one obtains 
\begin{eqnarray*}
\partial_{t}\tau_{2} & \approx & -\left\langle \boldsymbol{\theta}_{2}^{\dagger}\cdot\left({\cal B}_{1}-{\cal B}_{0}\right)\mathbf{u}_{2}\right\rangle _{1}\\
 & - & \left\langle \boldsymbol{\theta}_{2}^{\dagger}\left(t+T\right)\cdot\left({\cal C}_{1}-{\cal C}_{0}\right)\mathbf{u}_{2}\right\rangle _{1}\\
 & - & \left\langle \boldsymbol{\theta}_{2}^{\dagger}\cdot\left\{ \left({\cal B}_{2}-{\cal B}_{0}\right)\mathbf{u}_{1}\right\} \right\rangle _{2}\\
 & - & \left\langle \boldsymbol{\theta}_{2}^{\dagger}\left(t+T\right)\cdot\left({\cal C}_{2}-{\cal C}_{0}\right)\mathbf{u}_{1}\right\rangle _{2},
\end{eqnarray*}
where the first two terms in the right hand side containing the product
of two small quantities $\boldsymbol{\theta}_{2}^{\dagger}$ and \textbf{$\mathbf{u}_{2}$}
on the first interval $\left[-T_{0}/2,0\right]$ can be neglected.
Thus one obtains
\begin{eqnarray}
\partial_{t}\tau_{2} & \approx & -\left\langle \boldsymbol{\theta}_{2}^{\dagger}\cdot\left({\cal B}_{2}-{\cal B}_{0}\right)\mathbf{u}_{1}\right\rangle _{2}\nonumber \\
 & - & \left\langle \boldsymbol{\theta}_{2}^{\dagger}\left(t+T\right)\cdot\left({\cal C}_{2}-{\cal C}_{0}\right)\mathbf{u}_{1}\right\rangle _{2}.\label{eq:tau2_2}
\end{eqnarray}
Since $\mathbf{u}_{1}$ is the solution of Eq. (\ref{eq:general_form})
it satisfies the equation $-\partial_{t}\mathbf{u}_{1}+{\bf F}_{\omega_{0}}\left({\bf u}_{1}\right)+{\bf H}_{\omega_{0}}\left({\bf u}_{2}\right)=0$.
Using the relations (\ref{eq:aprox}) valid on the second interval
$\left[0,T_{0}/2\right]$ it can be rewritten on this interval in
the form

\begin{equation}
-\partial_{t}\mathbf{u}_{1}+{\cal B}_{0}\mathbf{u}_{1}+{\cal C}_{0}\mathbf{u}_{1}\left(t-T\right)\approx0.\label{eq:u1}
\end{equation}
The adjoint neutral mode $\boldsymbol{\theta}_{2}^{\dagger}$ satisfies
the equation
\begin{equation}
\partial_{t}\boldsymbol{\theta}_{2}^{\dagger}+\boldsymbol{\theta}_{2}^{\dagger}{\cal B}_{2}+\boldsymbol{\theta}_{2}^{\dagger}\left(t+T\right){\cal C}_{2}=0.\label{eq:theta2}
\end{equation}
Multiplying Eq. (\ref{eq:theta2}) by $\mathbf{u}_{1}$, subtracting
from the resulting equation $\boldsymbol{\theta}_{2}^{\dagger}$ multiplied
by Eq. (\ref{eq:u1}) and integrating over the second interval $\left[0,T_{0}/2\right]$
yields
\begin{gather*}
\left\langle \boldsymbol{\theta}_{2}^{\dagger}\cdot\left({\cal B}_{2}-{\cal B}_{0}\right)\mathbf{u}_{1}\right\rangle _{2}\approx-\left\langle \partial_{t}\boldsymbol{\theta}_{2}^{\dagger}\cdot\mathbf{u}_{1}+\boldsymbol{\theta}_{2}^{\dagger}\cdot\partial_{t}\mathbf{u}_{1}\right\rangle _{2}\\
-\left\langle \boldsymbol{\theta}_{2}^{\dagger}\left(t+T\right){\cal C}_{2}\cdot\mathbf{u}_{1}-\boldsymbol{\theta}_{2}^{\dagger}\cdot{\cal C}_{0}\mathbf{u}_{1}\left(t-T\right)\right\rangle _{2}.
\end{gather*}
Substituting this relation into (\ref{eq:tau2_2}) gives
\begin{gather*}
\partial_{t}\tau_{2}\approx\left\langle \partial_{t}\boldsymbol{\theta}_{2}^{\dagger}\cdot\mathbf{u}_{1}+\boldsymbol{\theta}_{2}^{\dagger}\cdot\partial_{t}\mathbf{u}_{1}\right\rangle _{2}+\left\langle \boldsymbol{\theta}_{2}^{\dagger}\left(t+T\right){\cal C}_{2}\cdot\mathbf{u}_{1}\right.\\
-\left.\boldsymbol{\theta}_{2}^{\dagger}\cdot{\cal C}_{0}\mathbf{u}_{1}\left(t-T\right)_{2}-\boldsymbol{\theta}_{2}^{\dagger}\left(t+T\right)\cdot\left({\cal C}_{2}-{\cal C}_{0}\right)\mathbf{u}_{1}\right\rangle _{2}.
\end{gather*}
Finally and integrating the full derivative $\partial_{t}\left(\boldsymbol{\theta}_{2}^{\dagger}\cdot\mathbf{u}_{1}\right)$
over the interval $\left[0,T_{0}/2\right]$ leads to 
\begin{eqnarray}
\partial_{t}\tau_{2} & \approx & \boldsymbol{\theta}_{2}^{\dagger}\left(T_{0}/2\right)\mathbf{u}_{1}\left(T_{0}/2\right)-\boldsymbol{\theta}_{2}^{\dagger}\left(0\right)\mathbf{u}_{1}\left(0\right)\nonumber \\
 & + & \left\langle \boldsymbol{\theta}_{2}^{\dagger}\left(t+T\right)\cdot{\cal C}_{0}\mathbf{u}_{1}-\boldsymbol{\theta}_{2}^{\dagger}{\cal C}_{0}\cdot\mathbf{u}_{1}\left(t-T\right)\right\rangle _{2}.\label{eq:tau2}
\end{eqnarray}
Note that the only nonzero elements of the $4\times4$ matrix
${\cal C}_{0}$ are those within the $2\times2$ block with the elements
having the indices $j,k\le2$. Hence, the last term in Eq.~(\ref{eq:tau2})
\begin{gather}
\left\langle \boldsymbol{\theta}_{2}^{\dagger}\left(t+T\right)\cdot{\cal C}_{0}\mathbf{u}_{1}-\boldsymbol{\theta}_{2}^{\dagger}{\cal C}_{0}\cdot\mathbf{u}_{1}\left(t-T\right)\right\rangle _{2}\nonumber \\
=-\left(\int_{0}^{\delta}+\int_{T/2}^{T/2+\delta}\right)\left[\boldsymbol{\theta}_{2}^{\dagger}\left(t+T\right){\cal C}_{0}\mathbf{u}_{1}\right]dt\label{eq:integr}
\end{gather}
contains only the asymptotical
expressions for the field components, which are assumed to be small
and are neglected in this study. Therefore, we can drop the last term
in Eq.~(\ref{eq:tau2}).

The equation for slow evolution of $\tau_{1}$ is derived in a similar
way to Eq. (\ref{eq:tau2}):
\begin{gather}
\partial_{t}\tau_{1}\approx\boldsymbol{\theta}_{1}^{\dagger}\left(0\right)\mathbf{u}_{2}\left(0\right)-\boldsymbol{\theta}_{1}^{\dagger}\left(T_{0}/2\right)\mathbf{u}_{2}\left(T_{0}/2\right)\nonumber \\
+\left\langle \boldsymbol{\theta}_{1}^{\dagger}\left(t+T\right)\cdot{\cal C}_{0}\mathbf{u}_{2}-\boldsymbol{\theta}_{1}^{\dagger}{\cal C}_{0}\cdot\mathbf{u}_{2}\left(t-T\right)\right\rangle _{1}.\label{eq:tau1}
\end{gather}
Note, that the terms $\boldsymbol{\theta}_{2}^{\dagger}\left(T_{0}/2\right)\mathbf{u}_{1}\left(T_{0}/2\right)$
{[}$\boldsymbol{\theta}_{1}^{\dagger}\left(0\right)\mathbf{u}_{2}\left(0\right)${]}
can be neglected in (\ref{eq:tau2}) {[}(\ref{eq:tau1}){]} due to
the fast decay of the leading tail of the TCS solution and trailing
edge of the adjoint neutral mode. The remaining terms $\boldsymbol{\theta}_{2}^{\dagger}\left(0\right){\bf u}_{1}\left(0\right)$
{[}$\boldsymbol{\theta}_{1}^{\dagger}\left(T_{0}/2\right){\bf u}_{2}\left(T_{0}/2\right)${]}
entering the Eq. (\ref{eq:tau2}) {[}(\ref{eq:tau1}){]} have very
different magnitudes except for the case where the TCSs are close
to be equidistant in the cavity, $\Delta\tau=\tau_{2}-\tau_{1}\approx\tau_{0}/2$.
This means that except for this case the TCS interaction is strongly
asymmetric and does not satisfy Newton's third law \cite{camelin2016electrical,vladimirov2018effect,vladimirov2022short}.
Thus, keeping only the second terms in the right hand sides of Eqs.
(\ref{eq:tau2}) and (\ref{eq:tau1}) one gets for the time evolution
TCS time separation $\Delta\tau=\tau_{2}-\tau_{1}$: 
\begin{equation}
\partial_{t}\Delta\tau\approx\boldsymbol{\theta}_{1}^{\dagger}\left(T_{0}/2\right)\mathbf{u}_{2}\left(T_{0}/2\right)-\boldsymbol{\theta}_{2}^{\dagger}\left(0\right)\mathbf{u}_{1}\left(0\right).\label{eq:Int1}
\end{equation}
The equation for the slow evolution of the phase difference
$\Delta\phi=\phi_{2}-\phi_{1}$ can be derived in a similar way. This equation reads: 
\begin{equation}
\partial_{t}\Delta\phi\approx\boldsymbol{\varphi}_{1}^{\dagger}\left(T_{0}/2\right)\mathbf{u}_{2}\left(T_{0}/2\right)-\boldsymbol{\varphi}_{2}^{\dagger}\left(0\right)\mathbf{u}_{1}\left(0\right).\label{eq:Int2}
\end{equation}
\bibliographystyle{abbrv}

\end{document}